\documentstyle[ltwol,epsfig]{article}

\input epsf

\def\be{\begin{equation}}
\def\ee{\end{equation}}
\def\bea{\begin{eqnarray}}
\def\eea{\end{eqnarray}}

\def\ot{\omega_T^{}}
\def\mot{M_{\omega_T}}
\def\got{\Gamma_{\omega_T}}
\def\gwwz{\Gamma_{WWZ}}
\def\zl{Z_L^{}}

\begin{document}

\title{ DISCOVERY LIMITS FOR TECHNI-OMEGA PRODUCTION IN $e\gamma$
COLLISIONS }

\author{ STEPHEN GODFREY, PAT KALYNIAK }

\address{ Ottawa-Carleton Institute for Physics, 
Department of Physics, Carleton University, Ottawa CANADA, K1S 5B6 \\
E-mail: kalyniak@physics.carleton.ca,
godfrey@physics.carleton.ca}   

\author{ TAO HAN }

\address{ Department of Physics,
University of Wisconsin, Madison,  WI 53706\\
E-mail: than@pheno.physics.wisc.edu}  


\twocolumn[\maketitle\abstracts{ In a strongly-interacting electroweak 
sector with an isosinglet
vector state, such as the techni-omega, $\ot$,
the direct $ \omega_T Z \gamma $ coupling
implies that an $\ot$ can be produced by $Z \gamma$ fusion in $e
\gamma$ collisions. This is a unique feature for high energy $e^+e^-$ or
$e^-e^-$ colliders operating in an $e\gamma$ mode. We consider the
processes $e^- \gamma \to e^- Z\gamma$ and $e^- \gamma \to e^- W^+ W^-
Z$, both of which proceed via an intermediate $\omega_T$. 
We find that at a 1.5~TeV $e^+e^-$ linear 
collider operating in an $e\gamma$ mode with an integrated luminosity 
of 200 fb$^{-1}$, we can discover
an $\omega_T$ for a broad range of masses and widths.}]

\section{ Introduction }
The mechanism for electroweak symmetry breaking remains the most
prominent mystery in elementary particle physics. In the Standard Model 
(SM), a neutral scalar Higgs boson is expected with a mass ($m_H$) to be 
less than about 800 GeV. In the weakly coupled supersymmetric
extension of the SM, the lightest Higgs boson should be lighter than
about 140 GeV. Searching for the Higgs bosons has
been the primary goal for current and future collider 
experiments \cite{higgs}. However, if no light Higgs boson is 
found for $m_H< 800$ GeV, one would anticipate that the interactions 
among the longitudinal vector bosons become strong \cite{sews}.  
This is the case when strongly interacting dynamics 
is responsible for electroweak symmetry breaking, such as
in Technicolor models \cite{TC}.

Without knowing the underlying dynamics of the strongly-interacting 
electroweak sector (SEWS), it is instructive to
parametrize the physics with an effective theory 
for the possible low-lying resonant states.
This typically includes an isosinglet scalar meson $(H)$ and an
isotriplet vector meson $(\rho_T)$  \cite{bagger94}.  
However, in many dynamical electroweak symmetry breaking models there 
exist other resonant states such as an isosinglet vector ($\omega_T$),
and isotriplet vector ($a_{1_{T}}$) \cite{chivukula90,rr}.  In fact it has
been 
argued that to preserve good high energy behavior for strong
scattering amplitudes in a SEWS, it is 
necessary for all the above resonant states to coexist \cite{han96}.  
It is therefore wise to keep an open mind and to consider other 
characteristic resonant states when studying the physics of a SEWS at 
high energy colliders.

Among those heavy resonant states, the isosinglet vector state
$\omega_T$ has rather unique features.
Due to the isosinglet nature, it does not
have strong coupling to two gauge bosons.
It couples strongly to three longitudinal gauge bosons $W^+_LW^-_LZ^{}_L$
(or equivalently electroweak Goldstone bosons $w^+w^-z$) and 
electroweakly to $Z \gamma$, $Z Z$, and $W^+ W^-$.
It may mix with the $U(1)$ gauge boson $B$, depending
on its hypercharge assignment in the model. The signal for $\ot$
production was studied for $pp$ collisions at 40 TeV and 17 TeV
\cite{chivukula90,rr}. It appears to be difficult to observe the
$\ot$ signal at the LHC. On the other hand,
the direct $\omega_T Z_L^{} \gamma$ 
coupling implies that an $\ot$ can be effectively produced by 
$\zl\gamma$ fusion in $e\gamma$ collisions.  This is a unique feature 
for high energy $e^+ e^-$ or $e^- e^-$ colliders operating in an 
$e\gamma$ mode. In this paper we concentrate on the $\ot$ production
at $e\gamma$ linear colliders. We first describe a 
SEWS model in Sec.~II in terms of an effective Lagrangian involving 
$\ot$ interactions. We then present our results in Sec.~III
for the production and decay of $\omega_T$ in $e\gamma$ colliders.
We show that a high energy $e\gamma$ linear collider 
will have great potential to discover $\omega_T$ with a mass
of order 1 TeV. We conclude in Sec.~IV.
\vspace*{-1.8pt}   

\section{$\ot$ Interactions}
For an isosinglet vector, a Techniomega-like state $\ot$,
the leading strong interaction can be parameterized by
\begin{equation}
{\cal L}_{\rm strong} = \frac{g_\omega}{v\Lambda^2}\ 
\varepsilon_{\mu\nu\rho\sigma}\ \omega_T^\mu \ 
\partial^\nu{w^+} \partial^\rho{w^-} \partial^\sigma z
\label{otwww}
\end{equation}
where $v=246$~GeV is the scale of electroweak symmetry breaking and 
$\Lambda$ is the new physics scale at which the strong dynamics
sets in. The effective coupling $g_\omega$ is of strong coupling
strength, and is model-dependent. It governs the partial decay
width $\Gamma(\ot \to w^+w^-z) \equiv \gwwz$. To study the $\ot$ signal in
a model-independent way, we will take the physical 
partial width as an input parameter to give the value for the factor
$(g_\omega/v\Lambda^2)$.

The effective Lagrangian describing the electroweak 
interactions of the $\omega_T$ 
with the gauge bosons can be written as \cite{chivukula90}:
\bea
{\cal L}_{\rm e.w.} & = & \chi \varepsilon_{\mu\nu\rho\sigma} 
[ g' \hbox{Tr} ( \frac{\sigma^3}{2} {B}^{\rho\sigma} 
\{ \Sigma^{\dag} D^\nu \Sigma , \omega_T^\mu \} ) \nonumber \\
& &  - g \hbox{Tr} ( \frac{\vec{\sigma}}{2}\cdot \vec{W}^{\rho\sigma} 
\left\{  { \Sigma D^\nu \Sigma^{\dag} , \omega_T^\mu  } 
\right\} ) ]
\eea
where  the covariant derivative is defined by
\begin{equation}
D^\mu\Sigma  = \partial ^\mu\Sigma -
ig\ \vec{W}^\mu \cdot \frac{\vec{\sigma}}{2}\Sigma + 
ig'\ \Sigma B^\mu \frac{\sigma_3}{2}
\end{equation}
$g$ and $g'$ are the $SU(2)_L$ and $U(1)_Y$ coupling constants, 
$\Sigma$ is the non-linearly realized representation of the 
Goldstone boson fields and transforms like 
$\Sigma \to L\Sigma R^{\dag}$. 
In the Unitary gauge $\Sigma \to 1$. 
With these substitutions ${\cal L}_{e.w.}$ leads to
\begin{eqnarray}
{\cal L}_{\rm e.w.} & \sim &  
2i e^2\ \chi \varepsilon_{\mu\nu\rho\sigma}\  
\omega^\mu_T \left[ 
\frac{2}{\sin\theta_w\cos\theta_w} \partial^\rho 
\gamma^\sigma Z^\nu  \right. 
\nonumber \\
& &   \left. { + \left( { \frac{\cos^2\theta_w}{\sin^2\theta_w}
- \frac{\sin^2\theta_w}{\cos^2\theta_w} } \right) 
\partial^\rho Z^\sigma Z^\nu } \right.
 \\
& & \left. {  + \frac{1}{\sin^2\theta_w} (\partial^\rho W^{+\sigma}
W^{-\nu}
+
\partial^\rho W^{-\sigma} W^{+\nu})  } \right] + \cdots
\nonumber
\end{eqnarray} 
where $\theta_w$ is the electroweak mixing angle.
The first term  gives the $\omega_T Z\gamma$ vertex, of importance for
production of the $\ot$ in $e\gamma$ colliders.
It involves the unknown coupling parameter $\chi$, where $e^2\chi$ may be 
of electroweak strength. 
Similarly, the $\ot ZZ$ and  
$\ot W^+W^-$ vertices, corresponding to the second and third terms,
respectively, are proportional to $\chi$. Hence, we take the
partial width of the $\ot$ into these two body states,
$\Gamma_{2-body} = \Gamma(\ot \to Z\gamma) + \Gamma(\ot \to W^+ W^-)
+
\Gamma(\ot \to ZZ)$ as input, to determine this coupling.
%

\begin{figure}[tbh]
\centering
\leavevmode
\epsfxsize=3.0in\epsffile{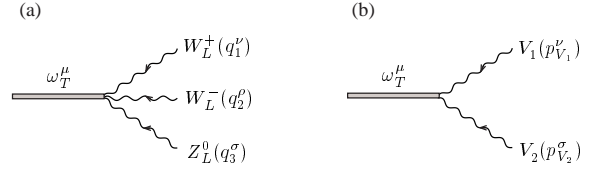}
\caption{ Effective interactions of $\ot$ with (a)
$w^+w^-z$ and (b) two vector bosons.}
\label{fig:vertex}
\end{figure}

The Feynman rules for the effective interactions of the $\ot$, represented
in 
Fig.~\ref{fig:vertex}, are given 
in Table ~\ref{Frules}. 


\begin{table}
\begin{center}
\caption{Feynman rules for the effective interactions of $\ot$}
\vspace{0.2cm}
\begin{tabular}{|l|l|}
\hline
\raisebox{0pt}[12pt][6pt]{Vertex} & 
\raisebox{0pt}[12pt][6pt]{Feynman rule} \\
\hline
\raisebox{0pt}[12pt][6pt]{$\ot w^+w^-z$} & 
\raisebox{0pt}[12pt][6pt]{$i\frac{g_{\ot}}{v\Lambda^2}
\epsilon_{\mu\nu\rho\sigma}
\epsilon^\mu(\omega)q^{\nu}_1 q^{\rho}_2 q^{\sigma}_3$}
\\
\hline
\raisebox{0pt}[12pt][6pt]{$\ot Z\gamma$} & 
\raisebox{0pt}[12pt][6pt]{$i\frac{4\chi e^2}{\sin\theta_w
\cos\theta_w}\epsilon_{\mu\nu\rho\sigma}\epsilon^\mu(\omega)
\epsilon^\nu(Z)p^\rho_\gamma\epsilon^\sigma(\gamma)$} \\
\hline
\raisebox{0pt}[12pt][6pt]{$\ot ZZ$} & 
\raisebox{0pt}[12pt][6pt]{$i4e^2\chi(\cot^2\theta_w
-\tan^2\theta_w)\epsilon_{\mu\nu\rho\sigma}\epsilon^\mu(\omega)
\epsilon^\nu_1p^{\rho}_{2}\epsilon^\sigma_2$} \\
\hline
\raisebox{0pt}[12pt][6pt]{$\ot W^+ W^-$} & 
\raisebox{0pt}[12pt][6pt]{$i\frac{2e^2\chi}
{\sin^2\theta_w}\epsilon_{\mu\nu\rho\sigma}\epsilon^\mu(\omega)
\epsilon^\nu_-(p^\rho_+-p^\rho_-)
\epsilon^\sigma_+$} \\
\hline
\end{tabular}
\end{center}
\label{Frules}
\end{table}
\vspace*{3pt}





\section{ Calculation and Results }
We consider the two signal processes which proceed via an intermediate 
$\omega_T$:
\begin{eqnarray}
\label{zgm}
e^- \gamma & \to & e^- \ot \to e^- Z \gamma \\
\label{wwz}
e^- \gamma & \to & e^- \ot \to e^- W^+ W^- Z .
\end{eqnarray}
Depending upon the  $\omega_T Z\gamma$ coupling, the signal 
cross section can be fairly large. The cross section expressions
are lengthy and we will not present them here. We choose to look at these
channels based on the distinctive signature of the first and the potential
enhancement of the second arising from its dependence on the strong
coupling $g_\omega$.

%
\begin{figure}[tbh]
\centering
\leavevmode
\epsfig{file=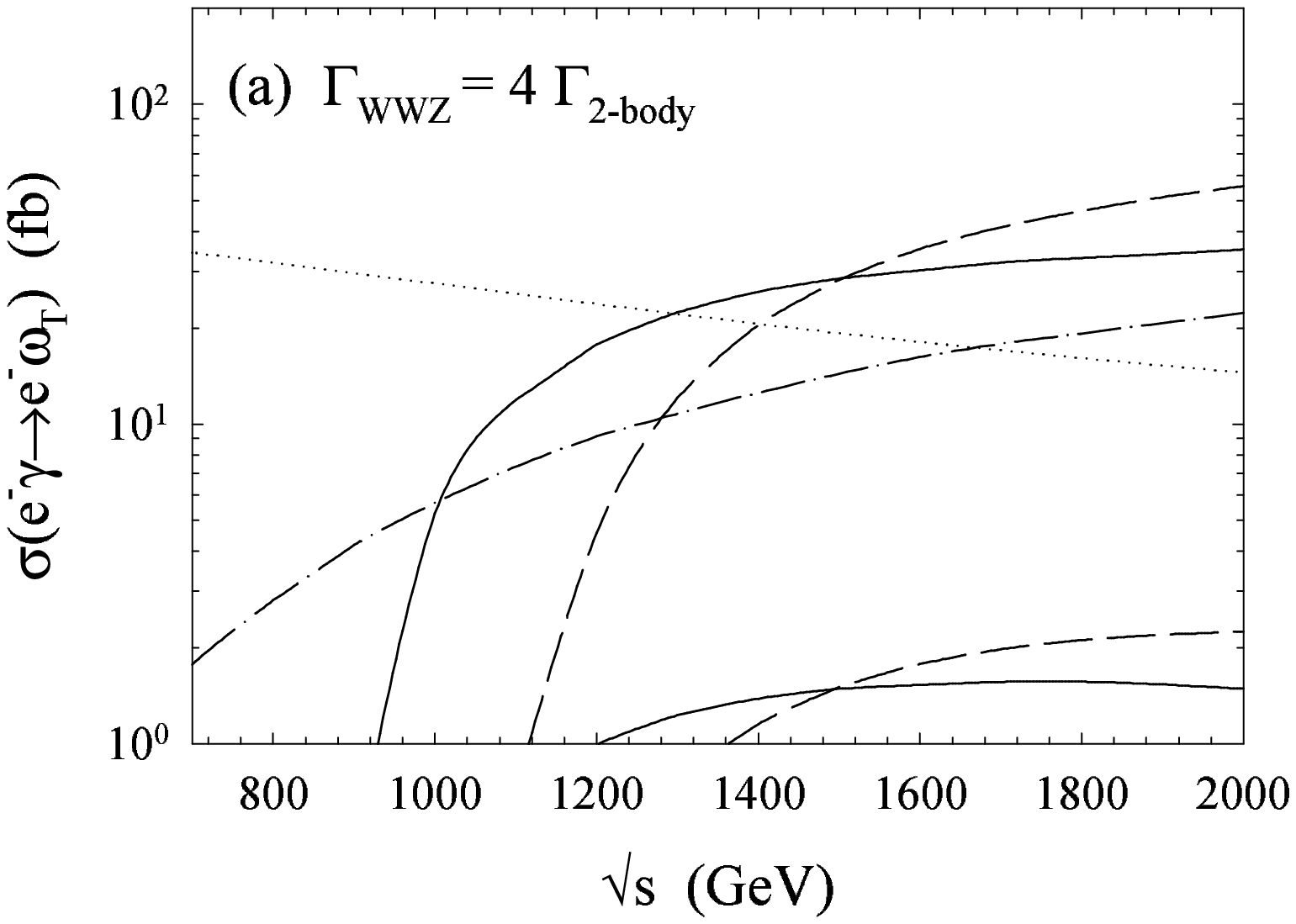,width=3.0in,clip=}
\epsfig{file=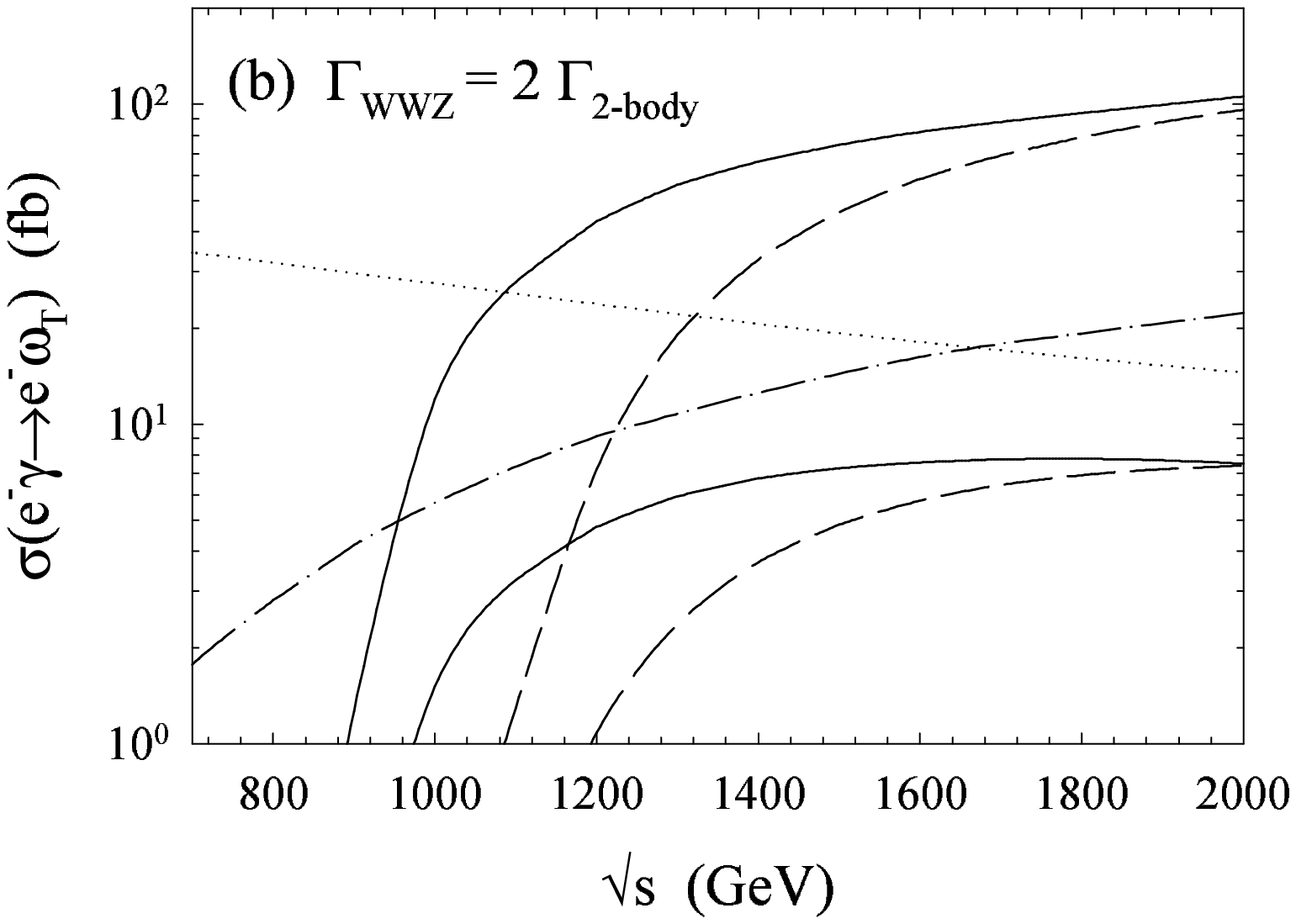,width=3.0in,clip=}
\caption[]{Cross sections versus the $e^+e^-$ c. m. energy
for the signal $e^-\gamma\to e^-\ot$ with
$\ot \to W^+W^-Z$ and $Z\gamma$ and the SM backgrounds.
The solid lines are for $M_{\omega_T}=0.8$~TeV 
and the long-dashed lines for $M_{\omega_T}=1.0$~TeV
In each case the curve with the larger 
cross section is for the $W^+W^-Z$ final state and the lower
for the $Z\gamma$ final state.  The dash-dot line is for 
the $e^-W^+W^-Z$ SM background and the dotted line is 
for the $e^-Z \gamma$ SM background.
In (a), $\gwwz=4\Gamma_{2-body}$ and in (b), 
$\gwwz=2\Gamma_{2-body}$. The choices of partial widths are given in
the text.}
\label{fig:xsecE}
\end{figure}

The SM background to the process $e\gamma \to e Z \gamma$ is the 
bremsstrahlung of photons and $Z$'s from the electron. The
background contribution to the $e^- W^+ W^- Z$ final state
has a complicated structure, mainly from the subprocesses
$e^+ e^- \to W^+ W^-, \gamma\gamma \to W^+ W^-$ with 
a radiated $Z$. We use the MADGRAPH package \cite{madg}
to evaluate the full SM amplitudes for the background processes.

In calculating the total cross sections for the $e Z \gamma$ signal 
and the
backgrounds, we impose the following ``basic cuts'' to roughly
simulate the detector coverage:
%
\bea
& & 170^\circ >\theta_e > 10^\circ,  \quad 165^\circ > \theta_\gamma >
15^\circ, 
\quad  \theta_{e\gamma} > 30^\circ,  \nonumber \\
& &  E_\gamma > 50\ {\rm GeV},
\label{Basic}
\eea
%
where $\theta_e$ is the polar angle with respect to the $e^-$ beam
direction in the lab ($e^+e^-$ c. m.) frame and $\theta_{e\gamma}$ is the 
angle between the outgoing $e^-$ and $\gamma$.
Only the cut on $\theta_e$ is relevant to the $e^- W^+ W^- Z$ 
process. The cuts on photons also regularize the infrared and
collinear divergences in the tree-level background calculations.

We have also implemented the back-scattered laser spectrum for
the photon beam \cite{Photon}. For simplicity, we have ignored 
the possible polarization for the electron and photon beams, 
although an appropriate choice of photon beam polarization may 
enhance the signal and suppress the backgrounds.

We present the total cross section for the signal and background
versus the $e^+e^-$ c. m. energy $\sqrt {s_{e^+e^-}}$ in 
Fig.~\ref{fig:xsecE} with various choices of $\ot$ mass and partial
widths. We have taken representative values for
$\mot$ of 0.8 (1.0)~TeV.
In Fig.~\ref{fig:xsecE}(a), we use partial widths 
$\Gamma_{2-body}=5$~(20)~GeV and $\gwwz=20$~(80)~GeV,
 setting $\gwwz=4\Gamma_{2-body}$. In Fig.~\ref{fig:xsecE}(b),
we
take $\Gamma_{2-body}=15$~(40)~GeV and $\gwwz=30$~(80)~GeV,
such that $\gwwz=2\Gamma_{2-body}$.
As noted above, the values for the couplings $\chi$ and $g_\omega$ are 
obtained using these partial widths as input. In Fig.~\ref{fig:xsecE}(b),
the 2-body decay modes represent a larger fraction of the total width and,
hence, the cross sections for the signal processes, which go via $Z\gamma$
fusion, are enhanced due to the larger value of $\chi$.
 We see that, for the parameters considered,  the signal cross sections
for the
$e^- W^+ W^- Z$ channel
are about $10-100$ fb once above the mass threshold and overtake the
background rates by as much as an order of magnitude. Such
high production rates imply that the linear collider would
have great potential to discover and study the $\ot$.
The cross sections for the $Z\gamma$ final state are lower
as expected and lie below the background with only the cuts of 
Eq.~(\ref{Basic}) imposed.

The reason that the cross section for $\mot=1.0$ TeV becomes
larger than that for 0.8 TeV in Fig.~\ref{fig:xsecE}(a) is because we have
chosen
relatively larger couplings for the 1.0 TeV $\ot$,
based on the input partial widths.

\begin{figure}[tb]
\centering
\leavevmode
\epsfig{file=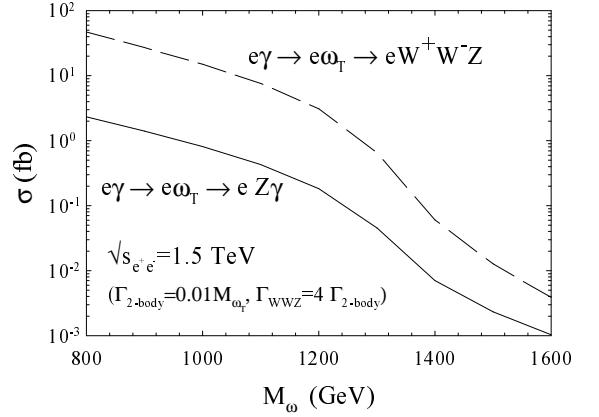,width=3.0in,clip=}
\caption[]{
Cross sections versus the $\mot$ 
for the signal $e^-\gamma \to e^-\ot$ for
$\sqrt{s_{e^+e^-}}=1.5$~TeV
with $\ot \to W^+W^-Z$ (dashed line) and $Z\gamma$ (solid line).}
\label{fig:xsecM}
\end{figure}

In Fig.~\ref{fig:xsecM}, we show the total signal cross sections
versus  $\mot$ for $\sqrt{s_{e^+e^-}}$=1.5~TeV. For simplicity,
the couplings are the values obtained by taking
$\Gamma_{2-body}=1\% M_{\omega_T}$ and 
$\gwwz=4\Gamma_{2-body}$. As expected, below the
mass threshold $\sqrt{s_{e\gamma}}<\mot$, the signal cross
section drops sharply. However, depending on the broadness
of the resonance, there is still non-zero signal cross section.

\begin{figure}[tb]
\centering
\leavevmode
\epsfig{file=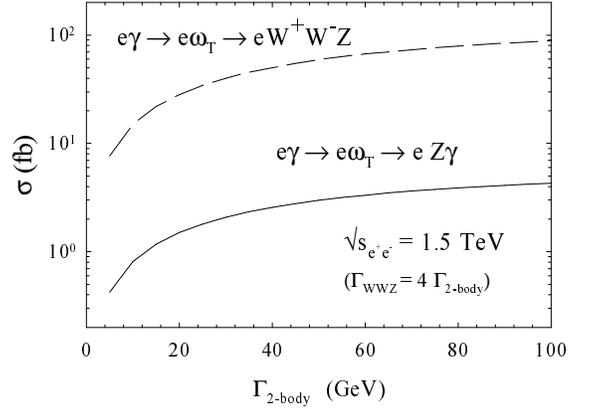,width=3.0in,clip=}
\caption[]{
Cross sections versus the partial width $\Gamma_{2-body}$
for the signal $e^-\gamma\to e^-\ot$ 
for $\sqrt{s_{e^+e^-}}=1.5$~TeV and $\mot=1$~TeV
with $\ot \to W^+W^-Z$ (dashed line) and $Z\gamma$ (solid line).}
\label{fig:xsecW}
\end{figure}

We have chosen partial decay widths of $\ot$ as input parameters
to characterize its coupling strength. It is informative to explore
how the cross section changes with the widths. Figure~\ref{fig:xsecW}
demonstrates this point, for $\sqrt{s_{e^+e^-}}$=1.5~TeV and 
$\mot$=1~TeV,
where we vary $\Gamma_{2-body}$ and take
$\gwwz=4\Gamma_{2-body}$. The signal cross section
rate and relative branching fractions for the two channels 
would reveal important information for the underlying SEWS
dynamics.

Although the SM backgrounds seem to be larger or, at best comparable, to
the signal
rate for the $Z\gamma$ channel,
the final state kinematics is very different between them.
Because the final state vector bosons in the signal are from the
decay of a very massive particle, they are generally very energetic
and fairly central. We thus impose further cuts to reduce the
backgrounds at little cost to the signal:
\begin{equation}
15^\circ < \theta_{\gamma,Z,W} < 165^\circ, \quad 
E_{\gamma,Z,W} > 150 \ {\rm GeV}.
\label{Cutii}
\end{equation}

\begin{figure}[tb]
\centering
\leavevmode
\epsfig{file=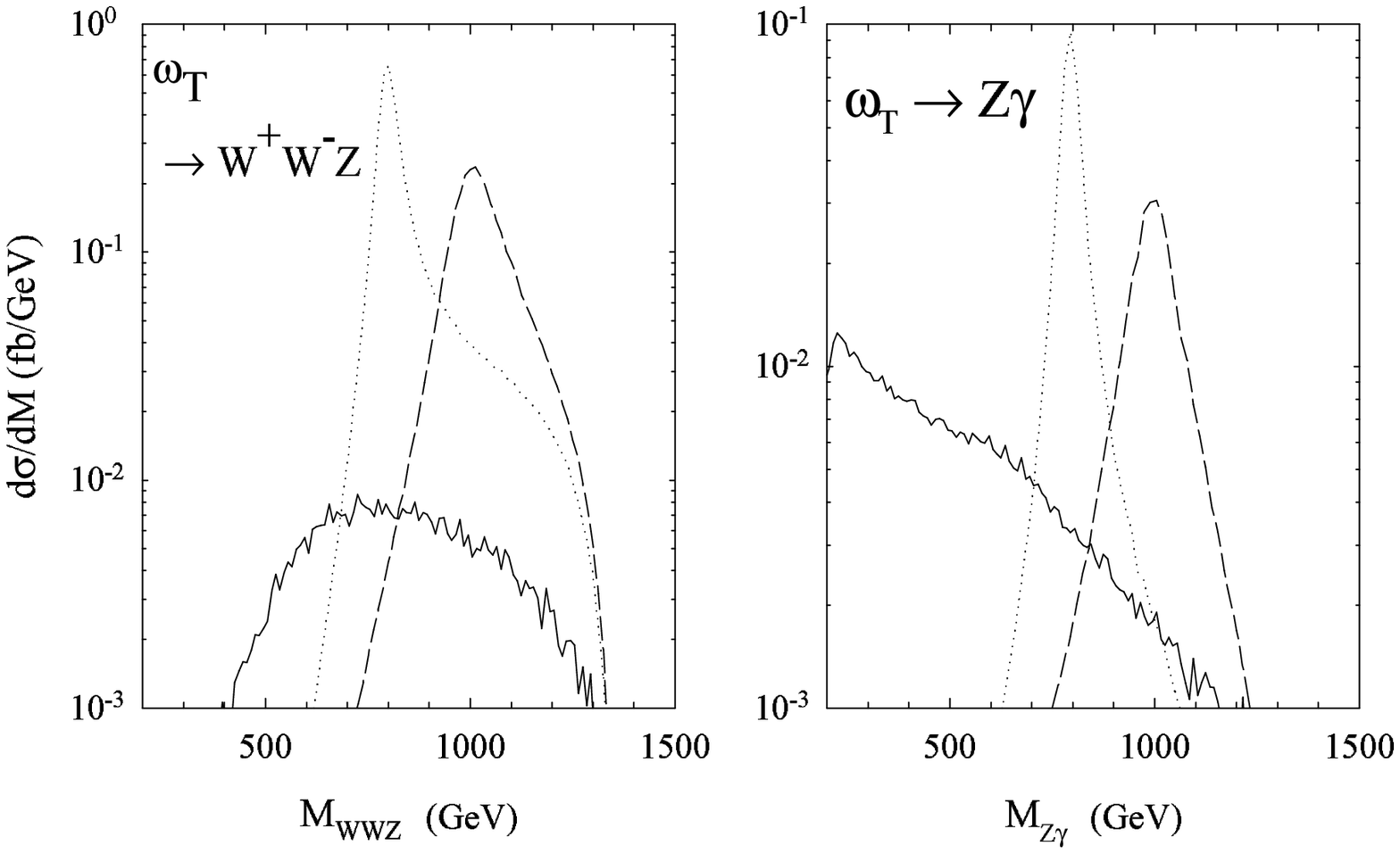,width=3.0in,clip=}
\caption[]{
Differential cross sections for $\sqrt{s_{e^+e^-}}=1.5$~TeV as a function of
the invariant mass of the $\ot$ decay products
$M(WWZ)$ and $M(Z\gamma)$
for the signal $e^-\gamma \to e^-\ot$ with
$\ot \to W^+W^-Z$ and $Z\gamma$.  The dotted lines are for 
${\mot}=0.8$~TeV ($\got=\Gamma_{2-body}+\gwwz=15+30$ GeV)
 and the dashed lines for ${\mot}=1.0$~TeV ($\got=40+80$ GeV).  
The SM 
backgrounds relevant to each case are given by the solid lines.}
\label{fig:Mass}
\end{figure}

The most distinctive feature for the signal is the resonance
in the invariant mass spectrum for $W^+W^-Z$ and $Z\gamma$ final
states. We demonstrate this in Fig.~\ref{fig:Mass} for both $W^+W^-Z$
and $Z\gamma$ modes. The cuts in both Eqs.~(\ref{Basic}) and
(\ref{Cutii}) 
are imposed. We see that a resonant structure at $\mot$ is evident 
and the SM backgrounds after cuts (\ref{Cutii}) are essentially
negligible for the particular choice of parameters shown.

\begin{figure}[tbh]
\centering
\leavevmode
\epsfig{file=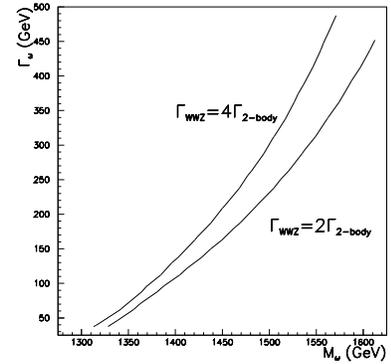,width=2.0in,clip=}
\caption[]{
Contours representing 10 signal events in the parameter
space for $\mot$ and $\got$ with $\sqrt{s_{e^+e^-}}=1.5$~TeV and an
integrated 
luminosity of 200 fb$^{-1}$.
Two choices of the ratio of
$\gwwz$ to $\Gamma_{2-body}$ are shown.}
\label{fig:contour}
\end{figure}

To further assess the discovery potential, we explore the 
parameter space for $\mot$ and $\got$ at a 1.5~TeV
linear collider. The signal for the $WWZ$ mode consists of both $W$'s
decaying hadronically and the $Z$ decaying hadronically or into electrons
or muons. The same $Z$ decay modes provide the signal for the $Z \gamma$
channel, along with the detected photon.  We assume an 80\%
detection efficiency for each of the $W$, $Z$, and $\gamma$ and
an integrated luminosity of 200 fb$^{-1}$. The contours given in
Fig.~\ref{fig:contour} represent 10 signal events. For these results, both
cuts (\ref{Basic}) and (\ref{Cutii}) are imposed. For the cases where the
cuts reduce the background to an insignificant level, this gives a
reasonable estimate of the discovery potential. We are undertaking a
more detailed analysis for that part of the parameter space where the
background is significant.

\section{ Conclusions }

A high energy $e\gamma$ collider is unique in producing an
isosinglet vector state such as $\ot$. We calculated the signal
cross sections for processes (\ref{zgm}) and (\ref{wwz})
in an effective Lagrangian framework. We found that signal rates
can be fairly large once above the $\mot$ threshold, although the
determining factor is the effective electroweak coupling of the $\ot$,
$\chi$. The signal characteristics are very different from the SM
backgrounds, making the discovery and further study of $\ot$ physics
very promising at the linear collider. With an integrated 
luminosity of 200 fb$^{-1}$ at $\sqrt{s_{e^+e^-}}=1.5$~TeV, 
one may discover an $\ot$ for a broad range of masses and widths.

\section*{Acknowledgements}
This research was supported in part by the Natural Sciences and Engineering 
Research Council of Canada, and in part by the U. S. Department of Energy
under Grant No. DE-FG02-95ER40896. Further support for T.H. was provided
by the University of Wisconsin Research Committee, 
with funds granted by the Wisconsin Alumni Research Foundation.

\end{document}